\documentstyle[12pt,aaspp4,tighten]{article}
\def\rosat{{\sl ROSAT~}}
\def\ha{H$\alpha$}
\newcommand\hii{H{\small II}}
\newcommand\sii{[\ion{S}{2}]}
\newcommand{\as}{$^{\prime\prime}~$}

\slugcomment{}
\begin{document}

\title{\bf Detection of X-ray-Emitting Hypernova Remnants in M101}

\author{Q. Daniel Wang}
\affil{Dearborn Observatory, Northwestern University}
\affil{ 2131 Sheridan Road, Evanston,~IL 60208-2900}
\affil{Electronic mail: wqd@nwu.edu}

\begin{abstract}
Based on an ultra deep (230~ks) \rosat\ HRI imaging of M101, 
we have detected 5 X-ray sources that coincide spatially with optical 
emission line features previously classified as supernova remnants in this 
nearby galaxy. Two of these coincidences (SNR MF83 and NGC5471B) most 
likely represent the true physical association of X-ray emission with 
shock-heated interstellar gas. MF83, with a radius of $\sim 134$~pc, is one 
of the largest remnants known. NGC5471B, with a radius of 
30~pc and a velocity of at least $350 {\rm~km~s^{-1}}$ (FWZI), is extremely 
bright in both radio and optical. The X-ray luminosities of these 
two shell-like remnants are $\sim 1$ and $3 \times 10^{38} {\rm~ergs~s^{-1}}$ 
(0.5-2~keV), about an order of magnitude 
brighter than the brightest supernova remnants known in our Galaxy and in
the Magellanic Clouds. The inferred blastwave
energy is $\sim 3 \times 10^{52}$~ergs for NGC5471B and 
$\sim 3 \times 10^{53}$~ergs for MF83. Therefore, the remnants likely 
originate in hypernovae, which are a factor of $\gtrsim 10$ more 
energetic than canonical supernovae and are postulated as being responsible
for $\gamma$-ray bursts observed at cosmological distances. The study of such
hypernova remnants in nearby galaxies has the potential to provide
important constraints on the progenitor type, rate, energetics, 
and beaming effect of $\gamma$-ray bursts.
\end{abstract}

\section{Introduction}

	The face-on grand-design spiral M101 (Fig. 1) is an ideal site 
for studying various X-ray source populations and their relationship to other 
galactic components. At a well-determined distance $D = 7.2\pm 0.4$~Mpc 
(thus $1^\prime = 2.1$~kpc; Stetson et al. 1998), 
the galaxy is probably the largest in the sky, except for members of our 
Local Group. The galaxy is also in a direction of exceptionally low 
foreground Galactic extinction [$E(B-V) \sim 0$; $N_H \sim 1.1 \times 
10^{20} {\rm~cm^{-2}}$]. A recent optical survey of the galaxy
by Matonick \& Fesen (1997, MF hereafter) has identified 93 supernova 
remnant (SNR) candidates. The galaxy also contains numerous giant \hii\
regions (GHRs) and complexes (GHCs), located primarily in outer spiral 
arms. Some of these GHRs are substantially brighter than the LMC 30 Dor 
nebula, which is the brightest \hii\ region in the Local Group.
Thus, M101 provides a particularly favorable circumstance to explore 
starburst-related phenomena.

	Based on an ultra-deep image from the \rosat\ High Resolution
Imager (RHRI), Wang et al. (1999) have detected 51 sources with X-ray
luminosities in the range of 0.4 - 20 $\times 10^{38} {\rm~ergs~s^{-1}}$
(0.5-2~keV). A cross-correlation of these X-ray sources with SNR 
candidates in the galaxy 
leads to a surprising discovery of 5 position coincidences. Unlike several
previously-known X-ray-bright SNRs (in other nearby galaxies), which result
from strong interaction between SN ejecta with dense circumstellar media, 
these apparently X-ray-emitting remnants in M101 are optically-resolved, 
and are thus relatively well evolved. Most importantly, the observed
X-ray emission, if arising in shock-heated interstellar gas, 
implies the blastwave energy of each remnant is a factor of $\gtrsim 10$ 
greater than the energy ($\sim 10^{51}$~ergs; e.g., 
Hughes et al. 1998) expected for an SN. 
Therefore, the remnants may well be the products of hypernovae, as have
been proposed to explain $\gamma$-ray bursts (GRBs) and their afterglows
(Paczy\'nski 1998; Fryer \& Woosley 1998). In this Letter, we 
describe this discovery and explore implications of such
remnants in the study of GRBs. The term ``hypernova'', proposed first by 
Paczy\'nski (1998) for a super-energetic GRB/afterglow event, is loosely 
used here to refer an explosion that appears too energetic to be a 
nominal SN.
  
\section{ Cross-Correlation between X-ray Sources and SNR Candidates}

	Table 1 lists 5 position coincidences between X-ray sources and
known SNR candidates in M101 (see also Fig. 1). Except for NGC5461B, 
which was discovered 
in radio continuum (Skillman 1985; Yang, Skillman, \& Sramek 1994) and 
was confirmed in optical  (Chu \& Kennicutt 1986; Chen \& Chu 1999), all 
other SNRs used in the cross-correlation are from 
the MF survey, which consists 
of both optical imaging and spectroscopic observations. The imaging 
observations covered five partially-overlapped regions of 
8\farcm7$\times$8\farcm7 each.
Emission nebulae with \sii/\ha\ $\gtrsim 0.45$ were identified as SNRs. 
Spectroscopic followup observations were carried out for 14 objects, including
MF83 in Table 1. The SNR positions (J2000; Table 1) are accurate to 1\as, 
radii $R$ to about $\pm 5$~pc, and both \ha\ Luminosities $L(H_\alpha)$
and \sii/\ha\ ratios to 15\% (MF). The 51 X-ray sources 
were detected with signal-to-noise ratios $> 3.5$ within a field of 
17$^\prime$ radius covered by the RHRI image (Wang et al. 1999). 
The RHRI was sensitive to photons in the 0.1-2~keV range and had 
a point spread function (PSF)
with a half power-encircled radius ranging from $\sim 2\farcs5$ on-axis 
to $20^{\prime\prime}$ at $17^\prime$ off-axis. We corrected the astrometry
of the image by matching centroids of 7 point-like X-ray sources 
with their optical counterparts. Table 1 includes only the SNRs whose 
position offsets $\Delta_{x-o}$ from X-ray source centroids are less 
than twice the $1\sigma$ uncertainty radii $r_{err}$ 
(Wang et al. 1999). Table 1 also 
includes the projected galactocentric radius $R_g$, expanding velocity 
$v_e$, and optical morphology as well as the 
X-ray source number and 0.5-2~keV luminosity of each remnant.

	What is the probability for a chance superposition? The MF survey
had a complete coverage of the central field of $ R_g \lesssim 8^\prime$ 
from the M101 nucleus, which is at R.A., Dec. (J2000) $ = 
54^h 3^m 12\fs7$, $54^\circ 20^\prime 54^{\prime\prime}$ (Israel, et al. 
1975). This field includes 29 X-ray sources and 87 SNRs. Our adopted
2$\sigma$ matching radii enclose a total region of 583~arcsec$^2$. The 
statistical probability to find by chance 1 match in the region, for 
example, is 0.45. Four SNRs in Table 1 are within the field. It 
is quite possible that one or two of the coincidences in Table 1 are
just the chance superposition. Indeed, H19 shows strong time variability
 and has a relatively hard spectrum (Wang et al. 1999). Thus the source 
most likely represents an accreting X-ray binary 
system. Although this system may still be 
related to the explosion responsible for the remnant, the X-ray emission 
is not due to shock-heated interstellar gas. To find 4 random SNR/X-ray 
source superpositions in the field, however, the probability 
falls to $\sim 10^{-3}$. Thus, we expect that at least some of 
the coincidences in Table 1 do represent physical associations
between the X-ray sources and SNRs.

	The most likely true X-ray-bright remnants in Table 1 are MF83 and 
NGC5471B. Their $\Delta_{x-o}$ are the smallest, well within the $1\sigma$ 
error radii. Both of them are shell-like (the other three have filled 
morphology; Table 1). Most importantly, these two SNRs are known to be the 
most unusual in M101 (MF; Yang, Skillman, \& Sramek 1994; Chen \& Chu 1999). 
MF83 is one of the largest SNRs known. Its high \sii/\ha\ 
ratio and bright [OIII] emission  clearly suggest that the optical 
nebula is shock heated; the [\ion{O}{3}]/H$_\beta (= 2.0$) ratio indicates a 
shock velocity of $\sim 86 {\rm~km~s^{-1}}$ (Table 17 in MF). 
The remnant, located in a region just off a 
spiral arm, shows no evidence for an association with 
any concentration of OB stars. NGC5471B is an extended, 
nonthermal radio source with a flux of 4.4$\pm0.8$ mJy at 30~cm and a 
spectral slope of --0.73 (Yang, Skillman, \& Sramek 1994). 
Based on optical echelle spectroscopy, Chu \& Kennicutt (1986)
detected shocked gas with velocity $\gtrsim 350 {\rm~km~s^{-1}}$ 
(FWZI) and inferred the required supernova blastwave energy as
$\sim 10^{52}$~ergs.
Therefore, the identification of both remnants are firm and they 
all appear to be unusually energetic.

\section{Constraints on Remnant Parameters}

	To characterize the spectral properties of X-ray-emitting plasma, 
we have further analyzed a \rosat\ PSPC observation of M101 (Wang et al.
1999). Though only
with a limited spatial resolution of $\sim 30^{\prime\prime}$ FWHM on-axis,
this observation provides a spectral resolution of 
$\delta E/E \sim 0.43 (0.93{\rm~keV}/E)^{0.5}$. The PSPC spectral data for 
NGC5471B is reasonably reliable, as it is located in a relatively isolated 
region, 22~kpc off the galaxy's nucleus.
We find that the Raymond \& Smith thermal plasma model provides a 
satisfactory fit to the data ($\chi^2/d.o.f. = 6.2/11$), 
giving the plasma temperature as 0.29(0.22-0.45) and 
the absorbing-gas column density as  $1.8(0.3-6.1) \times 10^{20} 
{\rm~cm^{-2}}$; the parameter intervals are all at the 90\% confidence. 
The solar metal abundance is assumed for X-ray-absorbing gas, which 
affects only the absolute value of the column density. 
The normalization of the model is 
\begin{equation}
K \equiv [10^{-14}/(4\pi D^2)]\int n_e^2 dV \sim 8.2 \times 10^{-5} 
{\rm~cm^{-5}}, 
\end{equation}
where $n_e$ is the electron density of the plasma, and the integration is 
over the entire volume of the remnant. The metal abundance $\zeta$ of the 
X-ray-emitting plasma is assumed as 13\% solar, as inferred from
accurate measurements of oxygen abundance in NGC5447 (Torres-Peimbert, 
Peimbert, \& Fierro,  1989).  The temperature of X-ray-emitting plasma 
indicates an expanding shock velocity of $\sim 5.0 \times 10^2 
{\rm~km~s^{-1}}$, consistent with the lower limit inferred from the 
optical kinematics. This, together with the relatively small optical size 
of NGC5471B, suggests the remnant is in the Sedov-Taylor phase. 

	We can thus further estimate physical parameters of NGC5471B, 
using the Sedov solution:
\begin{equation}
R = (20 {\rm~pc}) (2E_{52}t_4^2/n_0)^{1/5}, 
\end{equation}
where $R$, $E_{52}$, $t_4$, and $n_0$ are the 
radius (in units of pc), blastwave energy ($10^{52} {\rm~ergs}$), age 
($10^4$~yrs), and preshock
density (${\rm cm^{-3}}$) of the remnant. The shocked gas
temperature can be expressed as
\begin{equation}
 kT \sim (1.8 \times 10^{-3} {\rm~keV}) (R/t_4)^2, 
\end{equation}
and the postshock-to-preshock density ratio is $n_e/n_0 = 4$.
From (1), (2), and (3), we obtain
\begin{equation}
n_0 \sim 2 {\rm~cm^{-3}} (R/30 {\rm~pc})^{-3/2} (K/8\times 10^{-5} 
{\rm~cm^{-5}})^{1/2},
\end{equation}
\begin{equation}
E \sim (3 \times 10^{52} {\rm~ergs}) (R/30 {\rm~pc})^{3/2} 
(K/8\times 10^{-5} {\rm~cm^{-5}})^{1/2} (kT/0.3 {\rm~keV}), 
\end{equation}
and
\begin{equation}
t \sim (2 \times 10^4 {\rm~yrs})(kT/0.3 {\rm~keV})^{-1/2} (R/30 {\rm~pc}).
\end{equation}
This age is smaller than the time for the pressure-driven snowplow (PDS) 
phase to start (Cioffi, McKee, \& Bertschinger 1988),
$t_{PDS} \sim (5 \times 10^4{\rm~yrs})$$ n_0^{4/7}$$ E_{52}^{3/14} 
\zeta_{0.13}^{-5/14}$, consistent with our assumption of the Sedov-Taylor 
phase. We further estimate the total thermal energy as $\sim 10^{52}$~ergs 
and the gas cooling rate as $\sim 10^{39} {\rm~ergs~s^{-1}}$. The inferred
cooling time is $\sim 3 \times 10^5$~yrs, and thus the remnant is apparently 
not in a radiative phase. The estimate 
of $E$, in particular, is not sensitive to the uncertainty in the spectral 
parameters, because $K$ anti-correlates with $kT$ 
(Fig. 2). Thus, NGC5471B most likely originates in a 
super-energetic explosion --- a hypernova. 

	Similarly, we can combine the optical and X-ray results to 
constrain the physical properties of MF83. Its large radius and relatively 
low shock velocity suggest that this remnant has evolved into the PDS 
phase, which starts at the radius $R_{PDS} = (32 {\rm~pc}) 
E_{52}^{2/7} n_0^{-3/7} \zeta_{0.3}^{-1/7}$
and velocity $v_{PDS} = (376{\rm~km~s^{-1}}) E_{52}^{1/14} n_0^{1/7} 
\zeta_{0.3}^{3/14}$. $\zeta  = 0.3$ is assumed to be the same as the 
abundance measured for the nearby GHC NGC5461 
(Torres-Peimbert, Peimbert, \& Fierro,  1989), and the parameter values 
to be inferred are not sensitive to this assumption. The PDS evolution 
(Cioffi, McKee, \& Bertschinger 1988) follows 
\begin{equation}
R = R_{PDS}[(4/3)(t/t_{PDS})-1/3]^{3/10}
\end{equation}
and 
\begin{equation}
v_e = v_{PDS}[(4/3)(t/t_{PDS})-1/3]^{-7/10}.
\end{equation}
From these two equations and the values of $R$ and $v_e$  (Table 1), 
we get $E_{52} \sim 13 n_0^{1.2}$. If $n_0 \gtrsim 0.1
{\rm~cm^{-2}}$, MF83 is then a hypernova remnant (HNR) as well. 
We may estimate $n_0$ from the X-ray data on MF83. However, the 
remnant is not bright enough in X-ray and is in a relatively 
crowded region; no reliable X-ray spectral constraint is yet available. 
Nevertheless, if the X-ray spectrum (e.g., the plasma temperature) is not
too much different (a factor of $\lesssim 3$) from that of NGC5471B, the 
X-ray luminosity of MF83 suggests a total thermal energy 
$E_T \sim 6 \times 10^{52} {\rm~ergs}$. 
Taking into account the adiabatic energy loss during the PDS phase, 
$E \sim E_T (R/R_{PDS})^2$, 
we obtain $n_0 = (32/R)^{46/15} (376/v_e)^{11/5} E_{T,52}^{31/30} \sim 
2 {\rm~cm^{-3}}$; thus, $t \sim 1 \times 10^6$~yrs
and $E \sim 3 \times 10^{53} {\rm~ergs}$.

	While the above calculations are based on a simple SNR evolution
model, the consideration of potential complications may significantly alter the
estimates of the parameters. For example, if an explosion occurs in a 
pre-existing low-density (stellar wind) bubble, the actual blastwave 
energy can then be a factor of a few lower than our estimate 
(e.g., Hughes et al. 1998). The expanding velocity of MF86 is also uncertain. 
But we do not expect that such complications would
affect our results qualitatively. The strong X-ray emission as well as
the unusual optical and radio characteristics of NGC5471B and MF83 all
suggest that they originate in super-energetic explosions.

\section{Hypernova Remnants and GRBs}

	The detection of HNRs in nearby galaxies may have 
important implications for our understanding of GRBs.
The recent discovery of afterglows strongly suggests that GRBs
represent the most extreme high-energy events we have witnessed in the 
universe. In the extreme case (GRB 990123), the inferred isotropic energy
release ($\gtrsim 3 \times 10^{54}$~ergs) exceeds the rest mass of a neutron
star (Kulkarni et al. 1999), although the beaming of the emission 
may substantially reduce the energy requirement
(e.g., M\'esz\'aros, Rees, \& Wijers 1998 and references therein). 
Furthermore, such an event
may contain 10 times more energy in outflowing non-relativistic matter 
(Loeb \& Perna 1998), which is not observable during a GRB and its
afterglow. Most plausible models of  GRBs involve either 
the coalescence of two compact objects ---
two neutron stars or a neutron star and black hole
 --- or collapses of massive stars and/or 
 their mergers with compact companions (Paczy\'nski 1998; Fryer 
\& Woosley 1998). The later scenarios have 
become particularly attractive, because of the evidence that 
GRB/afterglow events appear close to star-forming regions. In all these 
models, an event leads to the formation of a black hole and provides an 
extractable energy of $\sim 10^{54}$~ergs (M\'esz\'aros, Rees, \& Wijers 
1998). This naturally raises the hope to find in nearby galaxies 
relics of similar events as they must have interacted with the interstellar
medium (ISM). Indeed, it has been suggested recently by Efremov, 
Elmegreen, \& Hodge (1998) and by Loeb \& Perna (1998) that some of 
\ion{H}{1} supershells or holes in the ISM may represent such relics. 
As demonstrated in these papers, however,
it is very difficult to distinguish the GRB scenario from other more 
conventional interpretations of the supershells: the combined action of 
stellar winds and SN of OB associations or impacts of high-velocity 
clouds. OB associations, for example, can be hard to detect 
within \ion{H}{1} supershells with typical ages $\gtrsim 10^7$~yrs. 

	In contrast, a relatively young HNR, or a young GRB relic, in a nearby 
galaxy can be identified relatively easily and can be studied in great 
details at multi-wavelengths. Except for its substantially greater
blastwave energy, an HNR resembles
an SNR and should have such distinct signatures as nonthermal radio continuum, 
high \sii/\ha\ ratio, large shock velocity, and strong soft X-ray 
emission, as observed from NGC5471B and MF83. These signatures are not 
expected for a supershell powered even by a young OB association. 
Energetically, one can argue that NGC5471B or MF83 might be a result of 
multiple SN explosions. But this would require an explosion rate of 
$\gtrsim 1$ per $10^4$~yrs, or a mean mechanical energy input of $\gtrsim 
10^{40} {\rm~ergs~s^{-1}}$, within the boundaries of each remnant.
Such concerted energy input could only occur in very massive OB 
associations, more massive than 30 Dor in the Large Magellanic Cloud
(e.g., Wang 1999). To be consistent with the sizes of NGC5471B and MF83, the
massive OB associations would also have to be very young (less than several 
times $10^6$~yrs). Such young OB associations, however, would
produce luminous \hii\ nebulae, similar to the 30 Dor nebula with 
predominantly thermal characteristics, which do not fit the observations
of either NGC5471B or MF83. One could further 
recognize the associations in optical images. But, there is no evidence for 
their presence within both NGC5471B and MF83 (Chu 1999, privately 
communications). Similarly, one can distinguish young HNRs from 
small ring-like or half-shell features which might be produced by 
recent impacts of high-velocity clouds. At an expected 
velocity comparable to the galactic disk rotation, the impact could hardly 
heat gas to a temperature of a few times $10^6$~K,
which should be typical for young HNRs.  
In conclusion, one should be able to identify relics of GRBs
if they are present as young HNRs in nearby galaxies.

	Could NGC5471B and MF83 be the relics of GRBs statistically? 
From our sample of 2 HNRs with their 
ages within $\sim 1 \times 10^6$~yrs, we may infer a rate of 
order 2 hypernovae per $10^6$~yrs in M101. Within the obvious uncertainties,
this rate agrees with theoretical estimates from stellar evolution models for
binary neutron star mergers or induced neutron star collapses in binary 
systems for an $L^*$ galaxy (e.g. Qin et al. 1998), and may also be 
comparable to estimated hypernova rates, which 
depend largely on the history of star formation (e.g., Fryer \& Woosley 
1998). Observationally, GRBs appear to be about $10^4 - 10^6$ rarer 
than SN (Wijers et al. 1998). Assuming an SN rate of 1 per 50~yrs 
(MF) in M101, one then expects 0.02-2 GRBs per $10^6$~yrs. Our inferred rate 
is at the upper end of this range. This may be explained by 
two important factors. 
First, the SN rate in M101 is of great uncertainty; the MF survey covers only 
part of the galaxy and a substantial number of SNRs as expected in GHCs are 
not detected (MF). The true SN rate may be considerably higher in M101.
Second, only a fraction of hypernovae are likely
observable as GRBs at cosmological distances, presumably due to a moderate 
beaming effect of $\gamma$-ray emission as expected in GRB models. Therefore,
the detection of the 2 remnants in M101 is reasonably consistent with the 
GRB statistics.

	Clearly, our detection of the 2 apparently young HNRs in M101 
represents only the first step to firmly establish their identity. No 
matter what this identity might be, however, these remarkable remnants must 
originate in very energetic events. Other nearby galaxies may also contain
similar remnants. In particular, there is a population of 
persistent X-ray sources with luminosities greater than the 
Eddington limit for a $\sim 1.6 M_\odot$ neutron star. Such superluminous 
sources tend to be associated with galaxies of active massive star 
formation, and are absent in members of the Local Group 
(e.g., Wang et al. 1999). It is conceivable that some of these sources, 
as in M101, may arise in HNRs. The upcoming {\it AXAF} and {\it XMM} 
will be able to spatially resolve such remnants and 
to obtain high spectral resolution data 
over the broad 0.1-10~keV range. The X-ray data, combined with detailed 
observations of the remnants in other wavelength bands, will provide
important constraints on the rate, blastwave energy distribution, 
and progenitor type of the events. Most excitingly, such a study may 
open up a new avenue to explore the origin and physics of GRBs.

\begin{acknowledgements}
	The author is indebted to Y.-H. Chu for her sharing insights and 
valuable information about the remnants during the course of this work,
which is supported partly by NASA LTSA grant NAG5-6413. He is also 
grateful to A. Loeb, S. Oey, and B. Paczy\'nski for stimulating discussions 
on GRBs as well as on hypernovae and their remnants. 
\end{acknowledgements}

{\hoffset=-1.2in
\begin{table}
\begin{center}
\caption{X-ray-emitting Remnants Candidates in M101}
{\scriptsize 
\begin{tabular}{lcccccccccccc}
\hline
\hline
\noalign{\smallskip}
SNR &R.A. &Dec. &$R_g$&$\Delta_{x-o}$&$r_{err}$(1$\sigma$) &L(\ha)&[SII] &Size$^a$&$v_e$&Optical &RHRI	&$L_x$	\\
Name&(h~~m~~s) &($^\circ~~^{\prime}~~^{\prime\prime}$)&(kpc)& ($^{\prime\prime}$)&($^{\prime\prime}$)&(${\rm ergs~s^{-1}}$)&/\ha&(pc)&(${\rm km~s^{-1}}$)&Shape &Source	&(${\rm ergs~s^{-1}}$)\\
\hline
MF37 	& 14 3 04.1 & 54 27 36 &14 &3.3 & 3.0 &4.3E37&0.72	&---	&---&---     &H19     &6.9E38 \\
MF54 	& 14 3 20.7 & 54 19 42 &3.6&6.2 & 3.1 &3.5E37&0.52	&27	&---&filled &H29     &1.4E38 \\
MF57 	& 14 3 24.2 & 54 19 44 &4.2&4.0 & 3.1 &1.1E37&0.47	&40	&---&filled &H30     &1.1E38 \\
MF83 	& 14 3 35.9 & 54 19 24 &7.7&1.8 & 3.1 &1.7E38&0.76	&267	&$\sim 86$ &shell   &H36     &1.2E38 \\
NGC5471B& 14 4 29.1 & 54 23 53 &24&3.1&3.4&4.0E38$^b$&0.51$^b$	&60$^b$	&$\gtrsim 175^c$&shell	 &H49     &3.0E38 \\
\noalign{\smallskip}
\hline
\noalign{\smallskip}
\end{tabular}}
\end{center}
$^a$ The sizes of MF SNRs are recalculated, using the M101 distance of 7.2~Mpc 
instead of 5.4~Mpc assumed by MF97.\par
$^b$ from Chen \& Chu (1999).\par
$^c$ from Chu \& Kennicutt (1986).
\end{table}}

\clearpage
\begin{figure} \caption{RHRI intensity contour map overlaid on an optical 
image of M101 from Palomar Sky Survey. The map is adaptively smoothed 
with a Gaussian of size adjusted to achieve a constant signal-to-noise 
ratio of $\sim 6$. The contours are at 1.1, 1.6, 2.4, 5.3, and $12
\times 10^{-3} {\rm~counts~s^{-1}~arcmin^{-2}}$; the lowest 
level is about 3$\sigma$ above the local X-ray background. 
The locations of the X-ray-emitting remnant candidates (Table 1) are 
marked. 
\label{fig1}}
\end{figure}

\begin{figure} \caption{68\%, 90\%, and 99\% confidence contours of the
spectral parameters: $kT$ vs. normalization of the  Raymond \& 
Smith thermal plasma fit to the X-ray spectral data of NGC5471B. The cross
marks the best-fit parameter values.
\label{fig2}}
\end{figure}
\end{document}